\documentclass[]{aa} 

\usepackage{graphicx}
\usepackage{txfonts}
\begin{document}

   \title{V-type candidates and Vesta family asteroids in the Moving Objects VISTA (MOVIS) Catalogue}


   \author{J. Licandro\inst{1,2}
          \and
          M. Popescu\inst{3,4}
          \and
          D. Morate\inst{1,2}
          \and
          J. de Le\'on\inst{1,2}
          }

   \institute{Instituto de Astrof\'{\i}sica de Canarias (IAC), C\/V\'{\i}a L\'{a}ctea s\/n, 38205 La Laguna, Spain
              \and
              Departamento de Astrof\'{\i}sica, Universidad de La Laguna, 38206 La Laguna, Tenerife, Spain
              \and
              Astronomical Institute of the Romanian Academy, 5 Cu\c{t}itul de Argint, 040557 Bucharest, Romania
              \and
                Institut de M\'ecanique C\'eleste et de Calcul des \'Eph\'em\'erides (IMCCE) CNRS-UMR8028, 
              Observatoire de Paris, 77 avenue Denfert-Rochereau, 75014 Paris Cedex, France
                          }

   \date{Start working 01 May 2015}

 
  \abstract
   { Basaltic asteroids (spectrally classified as V-types) are believed to be fragments of large differentiated bodies. The majority of them are found in the inner part of the asteroid belt, and are current or past members of the Vesta family. Recently, some V-type asteroids have been discovered far from the Vesta family supporting the hypothesis of the presence of multiple basaltic asteroids in the early  solar system. The discovery of basaltic asteroids in the outer belt challenged the models of the radial extent and the variability of the temperature distribution in the early solar system.
   }
   {We aim to identify new basaltic V-type asteroids using near-infrared colors of $\sim$ 40000 asteroids observed by the VHS-VISTA survey and compiled in the MOVIS-C catalogue.  We also want to study their near-infrared colors and to study the near-infrared color distribution of the Vesta dynamical family.}
   { We performed a search in the MOVIS-C catalogue of all the asteroids with ($Y-J$) and ($J-Ks$) in the range $(Y-J) \geq 0.5$ and  $(J-Ks) \leq 0.3$,  associated with V-type asteroids, and studied their color distribution. We have also analyzed the near-infrared color distribution of 273 asteroid members of the Vesta family and compared them with the albedo and visible colors from WISE and SDSS data. We determined the fraction of V-type asteroids in the family.} 
   {We found 477 V-type candidates in MOVIS-C, 244 of them outside the Vesta dynamical family.  We identified 19 V-type asteroids beyond the 3:1 mean motion resonance, 6 of them in the outer main belt, and 16 V-types in the inner main belt with proper inclination $i_p \le 3.0\degr$, well below the inclination of the Vesta family. We computed that $\sim$ 85\% of the members of the Vesta dynamical family are V-type asteroids, and only 1-2\% are primitive class asteroids and unlikely members of the family.}
   {This work almost doubles the sample of basaltic asteroid candidates in regions outside the Vesta family. Spectroscopic studies in the near-infrared and dynamical studies are needed to confirm their basaltic composition and to determine their origin.}


 \titlerunning{V-type candidates and Vesta family asteroids in MOVIS catalogue}
 
  \maketitle
%

\section{Introduction}

Basaltic asteroids are believed to be fragments of large bodies whose interiors reached the melting temperature of silicate rocks and subsequently differentiated \citep{2002aste.conf..183G}, forming a core of heavy minerals and a mantle of lighter minerals (olivine and pyroxene). Differentiation is a fundamental process that determines the later geological evolution of the objects,  so determining wether specific asteroids are differentiated or not is of scientific interest for understanding the process that sculpted the asteroid belt. Taxonomically, basaltic asteroids are classified as V-types, named after their most representative member (4) Vesta, which was for a long time the only known asteroid presenting a basaltic crust \citep{1970Sci...168.1445M,1993Sci...260..186B}. The NASA Dawn mission provided a detailed study of Vesta's composition (see  \citealt{2015Icar..259....1M} and references therein). The visible to near-infrared (VNIR, 0.4 - 2.5 $\mu$m) spectra of basaltic asteroids are characterized by two deep absorption bands around 1 and 2 $\mu$m, associated with the presence of pyroxene on their surfaces. This type of spectra are similar to those of Howardite-Eucrite-Diogenite meteorites (known as HEDs), which are thought to originate from V-type asteroids.

The majority of basaltic asteroids are found in the inner part of the asteroid main belt, between 2.15 and 2.5 AU. Most of them are current or past members of the largest asteroid family in the belt, the Vesta family, dynamically defined as having members with proper semi-major axis ($a_p$), eccentricity ($e_p$), and inclination ($i_p$) in the ranges $2.26\leq a_p \leq 2.48, 0.075 \leq e_p \leq 0.122$, and $5.6^{\circ} \leq i_p \leq 7.9^{\circ}$, respectively \citep{2015aste.book..297N}. The Vesta collisional family  originated from a collisional event that excavated a  crater in the surface of Vesta \citep{1997M&PSA..32S...9A,1997Icar..128...88T} and that occurred more than 1.2 Gyr ago \citep{2005A&A...441..819C}. Consistently, spectral studies show that a large fraction of the Vesta family members present a basaltic composition  \citep{1993Sci...260..186B,2001M&PS...36..761B}.

 Vesta and its dynamical family are not the only asteroids with a basaltic composition. In the last decades, more V-type asteroids were discovered far from Vesta family in the outer main belt (2.8 - 3.3 AU), as is the case of asteroid (1459) Magnya \citep{2000Sci...288.2033L}, or in the middle main belt (2.5 - 2.8 AU) \citep{2008Icar..198...77M,2009P&SS...57..229D}. The orbits of these asteroids suggest that they are not scattered Vesta family objects because  they cannot be explained by the typical ejection velocities produced during the cratering event. The case of (1459) Magnya is the most intriguing one because it is so far away from the Vesta family that it is almost impossible for it to be a fragment of the crust of Vesta. \cite{2002Icar..158..343M} proposed that Magnya is a fragment of another large differentiated asteroid that existed in the outer belt region. 

The presence of multiple differentiated parent bodies is also supported  by the meteorite record. \cite{2009Sci...325.1525B} report on the fall of a basaltic meteorite that has orbital properties and an oxygen isotope composition that suggest a parent body distinct from Vesta.  

The  existence of V-type asteroids outside the Vesta family is currently explained by the presence of multiple basaltic asteroids in the early solar system; this hypothesis is supported by the study of HED meteorites. \cite{2014MNRAS.444.2985H}  found at least three local sources of V-type asteroids, possibly associated with the parent bodies (349) Dembowska, (221) Eos, and (1459) Magnya. The discovery of basaltic asteroids in the outer main belt challenged the models of the radial extent and the variability of the temperature distribution in the early solar system, which generally did not predict melting temperatures in this region \citep{2014Icar..242..269H}. 

Using wide field surveys that cover huge fractions of the sky with several filters in the visible and near-infrared it is possible to search for V-type candidates. \cite{2006Icar..183..411R} presented a systematic method for identifying possible basaltic (V-type) asteroids using the 3rd release of the Moving Objects Catalog (MOC) of the Sloan Digital Sky Survey (SDSS). The method is based on the Principal Components Analysis of the MOC colors in the visible, combined with some refined criteria of segregation of the taxonomical classes. They found 505 asteroids exhibiting V-type colors, 263 of them outside the Vesta family, almost all of them in the inner asteroid belt, and 8 of them in the middle/outer belt. \cite{2008Icar..198...77M} also studied the distribution of V-type asteroids in the main belt using SDSS colors and dynamical criteria. They found 50 V-type candidates outside the Vesta family and using visible and near-infrared spectroscopy they confirmed that 10 of the 11 objects they observed are V-types. 

The aims of this paper are   to identify new V-type candidates using the near-infrared colors of $\sim$ 40000 asteroids observed by the VHS-VISTA survey and compiled in the MOVIS-C catalogue \citep{2016A&A...591A.115P} and  to study the near-infrared colors of all the Vesta family asteroids identified in the MOVIS-C catalogue. 

A general description of MOVIS-C catalogue and the criteria used to select V-type candidates is presented in Section 2. In Section 3 we determine the goodness of the V-type candidate selection by analyzing the candidates with known spectra, SDSS colors, and albedo. The distribution of the V-type candidates in the main belt and their near-infrared colors are analyzed in Section 4. The near-infrared colors of the Vesta family asteroids are studied in Section 5, and the conclusions are presented in Section 6.


\section {Identification of V-type asteroid candidates in MOVIS-C}

The Moving Objects VISTA Survey (MOVIS) is a compilation of the observations of known minor bodies of the solar system observed during the VISTA-VHS survey \citep{2016A&A...591A.115P}. The VISTA Hemisphere Survey (VHS) is an all sky survey that uses the 4.1m Visible and Infrared Survey Telescope for Astronomy (VISTA) wide field survey telescope located at ESO's Cerro Paranal Observatory in Chile,  equipped with a near-infrared camera with a 1.65$\degr$ field of view. VISTA-VHS uses the $Y$, $J$, $H$, and $Ks$ broadband filters and targets to image the entire southern hemisphere of the sky, covering about 19.000 square degrees. \cite{2016A&A...591A.115P} compiled the colors and the magnitudes of the minor planets observed by the VISTA survey in three catalogues: the detections catalogue,  MOVIS-D; the magnitudes catalogue,   MOVIS-M; and the colors catalogue,   MOVIS-C. The catalogues were built using the third data release of the survey (VISTA VHS-DR3). A total of 39.947 objects were detected. The colors found for asteroids with known spectral properties revealed well-defined patterns corresponding to different mineralogies. The distribution of MOVIS-C data in color-color plots shows clusters identified with different taxonomic types. All the diagrams that use in particular the $(Y-J)$ color separate the spectral classes much better than the common $(J-H)$ and $(H-Ks)$ colors used to date: even for quite large color errors ($\sim 0.1$)  the color-color plots $(Y-J)$ vs. $(Y-Ks)$ and $(Y-J)$ vs. $(J-Ks)$ clearly separate  the S- from the C-complex without overlapping between the regions.  The end members A-, D-, R-, and V-types also  occupy well-defined regions. 

\begin{figure}[h]
\begin{center}
\includegraphics[width=10cm]{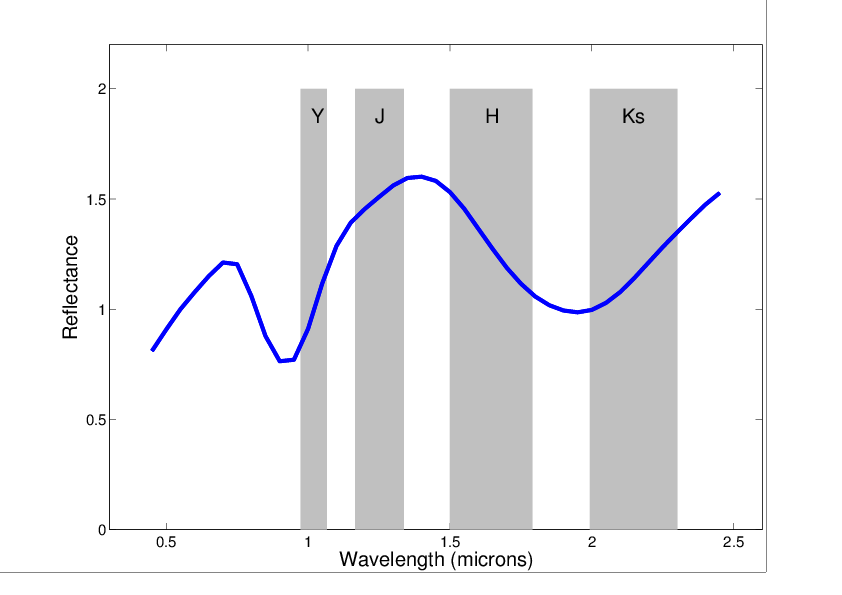}
\end{center}
\caption{Template spectrum of a V-type asteroid in the  Bus-DeMeo taxonomy. In gray the filter band-pass of the VISTA filters. The $(Y-J)$ and $(J-Ks)$ colors are strongly affected by the two deep absorption bands of the spectrum  (Band I centered at $\sim 1 \mu$m and Band II at $\sim 1 \mu$m)}.
\label{Vspectrum}
\end{figure}

Based on their distinctive spectral features asteroids with spectral properties similar to those of V-type asteroids can be easily identified using color-color diagrams obtained from observations with the $Y$, $J$, $H$, and $Ks$ filters \citep{2016A&A...591A.115P,2016AJ....151...98M}.  In \cite{2016A&A...591A.115P} we showed that  in the $(Y-J)$ vs. $(J-Ks)$ plot V-types appear as a separate group with $(Y-J) \geq 0.5$ and  $(J-Ks) \leq 0.3$ (see Fig. 15 in \citealt{2016A&A...591A.115P}).  In fact, the mean colors of asteroids with know taxonomic type in MOVIS-C show that the mean ($Y-J$) and ($J-Ks$) colors of V-types ($(\overline{Y-J}) = 0.64 \pm 0.09$ and $(\overline{J-Ks}) = 0.08 \pm 0.08$) are respectively 0.2 mag larger and more than 0.3 mag smaller than any other spectral class (see Table 1 in \citealt{2016A&A...591A.115P}). This is  explained well by the deep absorption bands around 1 and 2 $\mu$m (Band I and Band II) that characterize the  near-infrared spectra of V-types (see Fig. \ref{Vspectrum}) and the band-pass of the VISTA filters. We note that the $(Y-J)$ color is determined by the first absorption band and a large (red slope) value is indicative of a deep  1 $\mu$m band. On the other hand, the $(J-K_S)$ is also determined by the second absorption band and a low (blue slope) value is indicative of a deep  2 $\mu$m band. Based on the template spectrum for V-types from Bus-DeMeo taxonomy \citep{2009Icar..202..160D} we can determine that a V-type asteroid should have color values around $(Y-J) = 0.625 \pm 0.095$; $(Y-H) = 0.708 \pm 0.095$; $(Y-Ks) = 0.660 \pm 0.095$; $(J-H) = 0.082 \pm 0.095$; $(J-Ks) = 0.035 \pm 0.111$; $(H-Ks)=-0.047 \pm 0.050$, which are close to the middle of the region defined by the V-type candidates in the color-color  plots. The equivalent colors of the V-types were computed using the taxonomic template spectrum, the response curve of the VISTA filters, and the solar colors, as explained in \cite{2016A&A...591A.115P}.

In this work we use the colors obtained from the recently updated VHS-VISTA survey data release (VHSv20160114)  to define V-type candidates in the MOVIS-C catalogue: asteroids having $(Y-J) \geq 0.5$ and $(J-Ks) \leq 0.3$ with uncertainties in both colors $< 0.1$. These values correspond to the average values of the objects located close to the limit that separates V-types from the rest of the classes in the $(Y-J)$ vs. $(J-Ks)$ plot in Fig. 15 from \cite{2016A&A...591A.115P}. This plot shows a distinctive gap between this group and the rest of the objects. We base our selection criteria in these two colors because they clearly separate  the V-types from the other classes and because of observational constraints due to the particular observing procedure used in the VHS survey: (1) the number of objects with $K_s$ data is larger than those with $H$ data and (2) the $(J-Ks)$ colors are more accurate than the $(J-H)$ ones mainly because of the shorter time interval between the observations in both filters \citep{2016A&A...591A.115P}.

As discussed above, the existence of V-type asteroids outside the Vesta family can be  explained by the presence of multiple basaltic asteroids in the early solar system;  for the analysis of the V-type candidates, we will differentiate between those candidates that are members of the Vesta family according to \cite{2015aste.book..297N}  and those that are non-Vesta family members.

A total of 477 V-type candidates have been identified in the MOVIS-C catalogue using the criteria described above, 233 of them  members of the Vesta family (see Table \ref{VVesta}) and 244 non-Vesta family members (see Table \ref{VnoVesta}). Only 22 of our V-type candidates, marked with an asterisk in Tables \ref{VVesta} and \ref{VnoVesta}, are also identified as V-type candidates by \cite{2006Icar..183..411R} based on their SDSS colors.

\begin{table}[h]
\caption{List of MOVIS-C V-type candidates that belong to the Vesta collisional family and have color uncertainties $\leq$0.1. Objects marked with an asterisk are also identified as V-type candidates by \cite{2006Icar..183..411R} based on their visible colors.}
\centering
\begin{tabular}{l l l l l l l}\hline
1959 &15781 &29334 &43388$^{*}$ &62406 &96698 &\\
1979 &15881 &29979 &44097 &65068 &96890 &\\
2011 &16234 &29994 &44805 &65504 &98086$^{*}$ &\\
2508 &17139 &30097 &44877 &66111 &98167 &\\
3153 &17162 &30329 &45792 &67498 &99381 &\\
3613 &17431 &30358 &46281 &68759 &102813 &\\
4311 &17562 &30818 &46698$^{*}$ &68782 &103042 &\\
4444 &17769 &31132 &48644 &68801 &103132 &\\
4993 &17976 &31517 &48734 &68879 &107709 &\\
5051 &18508 &31575 &50048 &69742$^{*}$ &108250 &\\
5307 &18581 &31778 &50082 &70248 &111976 &\\
6014 &18754 &32276 &50084$^{*}$ &70277 &112087 &\\
6085 &19025 &32541 &50086 &70940 &117819 &\\
6096 &19573 &33049 &50241 &72246 &118295 &\\
6506 &19680 &33100 &50248 &73109 &118305 &\\
6877 &20252 &33477 &51368 &73174$^{*}$ &118532 &\\
7005 &20302 &33491 &51487 &74010 &119085 &\\
7012 &21633 &33512 &51628 &74898 &119136 &\\
7810 &21692 &33590 &51687 &74924 &122101 &\\
9076 &21883 &33875 &52792 &74936 &130756 &\\
9204 &22080 &34081 &53000 &75636 &131609 &\\
9616 &22155 &34534$^{*}$ &53580 &77122 &132010 &\\
10056 &22654 &35222 &53734 &77324 &132347 &\\
10614 &23522 &35284 &54084$^{*}$ &77590 &136439 &\\
11189 &24085 &35414 &54152 &80798 &137512 &\\
11326 &24115 &36021 &55315 &80800 &138192 &\\
11522 &24255 &36761 &56169 &84328 &147141 &\\
12340 &24261 &36834 &56599 &86520 &153392 &\\
12591 &25220 &37113 &57104 &88781 &161029 &\\
13054 &25708 &37149 &57818 &88958 &162518 &\\
13164 &26238 &37192 &57930$^{*}$ &90548 &170306 &\\
13191 &26611 &37234 &57943 &90639 &170309 &\\
13287 &27539 &38335 &58271 &91343 &172447 &\\
13530 &27939$^{*}$ &38732$^{*}$ &59215 &92646 &175841 &\\
13569 &28397 &38876 &59336 &92686 &178112 &\\
13855 &28543 &38879 &59569 &93339 &225334 &\\
13994 &28737 &39949$^{*}$ &61521 &93394$^{*}$ &243137 &\\
15032 &28902 &41558 &61741 &94355 &244232 &\\
15756 &29186 &42644 &61986 &94413 & &\\
\hline
\end{tabular}
\label{VVesta}     
\end{table}

\begin{table}[h]
\caption{List of MOVIS-C V-type candidates with color uncertainties $\leq$0.1 that do not belong to the Vesta family. The objects marked with an asterisk are also identified as V-type candidates by \cite{2006Icar..183..411R} based on their visible colors.}
\centering
\begin{tabular}{l l l l l l l}\hline
2275 &12172 &26842 &44878 &70081 &113910 &\\
2452 &12612 &27343 &44930 &71826 &114858 &\\
2486 &12787 &27373 &44953 &71850 &122008 &\\
2763 &12789 &27383 &45323 &71960 &122243 &\\
2888 &13194 &27770 &45893 &73076 &122267 &\\
3188 &13380 &28461 &45942 &73658 &123786 &\\
3331$^{*}$ &13398 &29384 &47459 &74596 &126981 &\\
3536 &13721 &29677 &47476 &74912 &130972 &\\
3882 &13760 &29733 &47837 &75080 &133202 &\\
3900 &15121 &30893$^{*}$ &48036 &75289 &135575 &\\
3954 &15506 &30961 &49214 &75661 &136229 &\\
4228 &15678 &31460 &50035 &77972 &145887 &\\
4693 &15846 &31599 &50091 &79426 &147703 &\\
5150 &16169 &31677 &50105 &80345 &152165 &\\
5328 &16352$^{*}$ &31775 &50215 &80463 &152933 &\\
5524 &16477 &32581 &51511 &82349 &159601 &\\
5631 &16605 &33366 &52132 &82455 &164235 &\\
5713 &16873 &33513 &52819 &86263 &165142$^{*}$ &\\
5758 &17001 &33562 &52985 &86284 &166973 &\\
5875 &17057 &33628 &52995 &86768 &177529 &\\
5952 &17546 &35062$^{*}$ &53608 &89270 &179851 &\\
6046 &17739 &36644 &54367 &89729 &180703 &\\
6363 &18012 &37386 &55456 &89729 &183004 &\\
6406 &18644 &37404 &55769 &90223 &188102 &\\
6442 &19257 &37730 &56369 &94680 &189231 &\\
6584 &19281 &38317 &56456 &96653 &197480 &\\
6587 &19294 &38744 &57278 &96823 &203379 &\\
6853 &19518 &39175 &57454 &98231 &207473 &\\
7223 &19619 &39465 &59423 &98654 &218144 &\\
7459 &19679 &39926 &59834 &99579 &235061 &\\
7675 &19738$^{*}$ &40258 &61354 &100772 &240551 &\\
7823 &19969 &40378 &61736 &101029 &245202 &\\
7998 &19983 &40521$^{*}$ &61985 &102601 &249866 &\\
8644 &20188 &40708 &63673 &102986 &284907 &\\
8921 &21307 &41433 &63708 &103105 &313008 &\\
9064 &22880 &41463 &65949 &103828 &322744 &\\
9147$^{*}$ &22892 &41776 &67477 &109080 &326769 &\\
9197 &24140 &41793 &67792 &111947 &354036 &\\
9495 &24604$^{*}$ &42656 &67876 &112839 &364694 &\\
9495 &25979 &43885 &68141 &112841 & &\\
9746 &26097 &44541 &69255 &113516 & &\\
\hline
\end{tabular}
\label{VnoVesta}     
\end{table}

\section{Visible spectra, colors, and albedo of MOVIS V-type candidates}

To see how efficiently  MOVIS-C can identify V-type candidates, we searched for published spectra of these objects in the two largest spectroscopic databases: the Small Main-Belt Asteroid Spectroscopic Survey (SMASS, \citealt{1995Icar..115....1X}, \citealt{2002Icar..158..106B}) and the Small Solar System Objects Spectroscopic Survey (S3OS2, \citealt{2004Icar..172..179L}). Nine of the candidates have spectra in the SMASS database, while there are none  in the S3OS2. We also searched for visible and/or near-infrared spectra in the literature and found that there are spectral data of a total 15 of our V-type candidates, 13 of which  already spectroscopically classified as V-types (see Table \ref{Vspectra}).  From those that are not, one is classified as Sr, and the other is ambiguously classified as V- or R-type. We note  that the spectra of R- or Sr-types are also characterized by deep absorption bands close to 1 and 2 $\mu$m.  On the other hand, eight objects with MOVIS-C data are already classified as V-types based on their visible spectrum obtained by SMASS survey. Six of these objects have well-determined $(Y-J)$ and $(J-Ks)$ colors: (2508) Alupka, (3498) Belton, (3536) Schleicher, (3900) Knezevic, (4311) Zguridi, and (4993) Cossard. Except for (3498) Belton, which has a $(J-Ks)$ error slightly larger than the 0.1 limit we used, all of them are identified as V-type candidates in this paper. So, 87\% of the V-type candidates identified in MOVIS-C with known spectra are V-type, and all the asteroids in MOVIS-C classified as V-type in the SDSS are identified as V-type candidates. This is indicative of the very high success rate of identifying V-types using MOVIS-C and the color criteria presented in this work.


\begin{table}[h]
\caption{\label{Vspectra}MOVIS-C V-type candidates with known spectral class.}
\centering
\begin{tabular}{l c c c}\hline
Object & Reference &Vesta family        &Class\\ \hline
2011 Veteraniya &4,11   &Y      &V \\
2486 Metsahovi  &1,6            &N      &V \\
2508 Alupka             &2              &Y      &V \\
2763 Jeans (*)          &2,5            &N      &V \\
3536 Schleiche  &2              &N      &V \\
3613 Kunlun             &12             &Y      &V \\
3900 Knezevic           &2              &N      &V \\
4311 Zguridi            &2              &Y      &V  \\
4993 Cossard            &2,4            &Y      &V \\
5051 Ralph              &2              &Y      &Sr \\
6406 Vanavara           &1,4            &N      &V \\
9147 Kourakuen  &3,10   &N      &V \\
10614 1997UH1   &3              &Y      &V or R \\
27343 Deannashea        &8              &N      &V \\
40521  (1999 RL95)      &7,9            &N      &V \\
\hline
\end{tabular}
\tablefoot{(1) \cite{2006A&A...459..969A}; (2) \cite{2002Icar..158..106B}; (3) \cite{2011A&A...533A..77D}; (4) \cite{2011MNRAS.412.2318D} ; (5) \cite{2004Icar..171..120D}; (6) \cite{2016MNRAS.455..584F}; (7) \cite{2008Icar..198...77M}; (8) \cite{2010Icar..208..773M};  (9) \cite{2008Icar..194..125R}; (10) \cite{2012A&A...544A.130P}; (11) \cite{1995Icar..115....1X}; (12) http://smass.mit.edu/catalog.php .}
\end{table}

We next looked at the SDSS colors and visible albedo ($p_V$) of our V-type candidates, confirming that most of them also have  visible colors and albedo compatible with V-type asteroids. The SDSS colors $(i-z)$ and $a^{*}$ (as defined by  \citealt{2001AJ....122.2749I}), and the $p_V$ calculated from the Wide Infrared Survey Explorer (WISE) data \citep{2011ApJ...741...68M} were retrieved using the MP$^3$C Minor Planet Physical Properties Catalogue\footnote{\url{http://mp3c.oca.eu/}}.

\begin{figure}[h]
\begin{center}
\includegraphics[width=8cm]{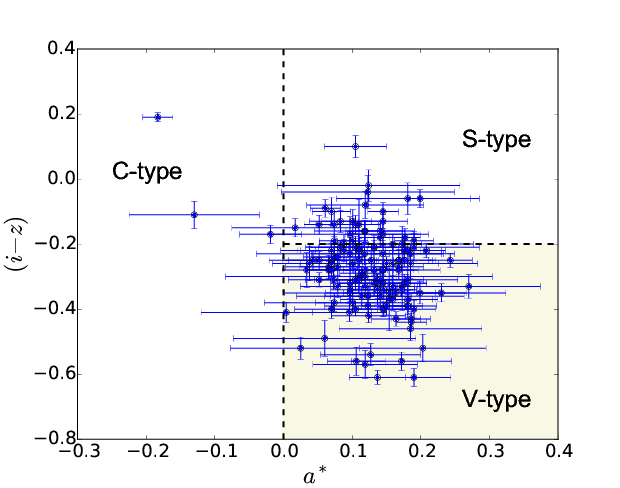}
\end{center}
\caption{ SDSS colors of the V-type candidates presented in Tables \ref{VVesta} and \ref{VnoVesta}. The great majority of them are in the region of the V-types according to \cite{2006Icar..183..411R}. }
\label{sdss}
\end{figure}

Using SDSS colors from the 3rd release of the MOC, \cite{2006Icar..183..411R} identified a few hundred  V-type candidates. They showed that V-type asteroids appear well segregated in the $(i-z)$ vs. $a^{*}$ color-color plot, forming a cluster in the region $a^{*} > 0$ and $(i-z) < -0.2$. A total of 144 of our V-type candidates presented SDSS colors \citep{2002AJ....124.2943I} in the 4th release of the MOC. Following this criterion (see Fig.~\ref{sdss}), 111 out of the 144 V-type candidates presented in this paper (77\%) have SDSS colors compatible with V-type asteroids. We note  that only 22 of our 144 V-type candidates are identified as V-types in \cite{2006Icar..183..411R}, likely because they used the 3rd release of the MOC that have photometric data for 204,305 moving objects, while the 4th MOV release used in this paper have data for 471,569 moving objects.
Considering the error bars of the SDSS colors only 10 asteroids present visible colors far from the V-type cluster ($a < 0.1$ and $(i-z) < -0.1$). Only 2 of the 144 present SDSS colors that place them in the region associated with the primitive C-complex ($a^{*}< 0.0$), namely (166973) 2003 OV$_5$ and (15032) Alexlevin.

\begin{figure}[h]
\begin{center}
\includegraphics[width=8cm]{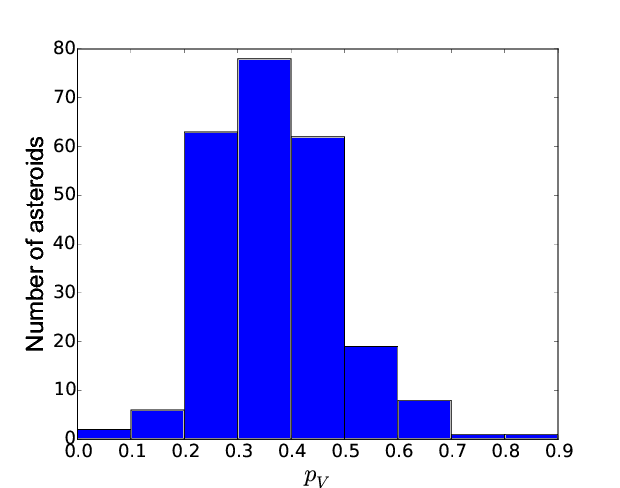}
\end{center}
\caption{Visible albedo ($p_V$) distribution of the V-type candidates in Tables \ref{VVesta} and \ref{VnoVesta}. The majority of the objects have large albedos ($p_V > 0.2$), compatible with V-type asteroids.}
\label{Valbedo}
\end{figure}

A total of 240 of our V-type candidates have visible albedo ($p_V$) obtained from WISE observations  \citep{2011ApJ...741...68M}. The albedo distribution (see Fig. \ref{Valbedo}) is typical of that of rocky asteroids, with $(\overline{p_V}) = 0.37 \pm 0.12$, with 232 objects (87\%) with $p_V > 0.2$ and only 3 with $p_V < 0.15$ (compatible with C-class objects within the uncertainties): (166973) 2003 OV$_5$, (5524) Lecacheux, and (112839) 1998 SA$_{111}$. Low-albedo asteroids cannot be basaltic, so the selection criteria includes $\sim$ 1.5\% of asteroids that are clearly not V-type likely owing to errors in the color determination.

We note  that (166973) 2003 OV$_5$ has both SDSS colors and $p_V$ compatible with a primitive class asteroid. On the other hand, the albedo of (15032) Alexlevin is too high ($p_V = 0.29$) and (5524) Lecacheux and (112839) 1998 SA$_{111}$  have $a^{*} >$ 0 ($0.132 \pm 0.016$ and $0.105 \pm 0.034$, respectively).

Finally, of the 477 candidates, 76 have both SDSS colors and WISE albedo, and 56  of these 76 (74\%) have the albedo and SDSS colors of the V-types ($p_V > 0.20$, $a^{*} > 0$ and $(i-z) < -0.2$). We note that the $74\%$ success rate of detecting objects that are V-types according to MOVIS, SDSS, and WISE data depends on the success rate (and uncertainties) of all the surveys. The success rate of MOVIS alone of detecting V-type candidates among those that have been already classified because of their visible or near-infrared spectrum (confirmed V-types) is higher, $\sim 87 \%$ (13 of 15 objects).
 


\section {V-type candidates outside the Vesta family}

\subsection{Distribution of V-type candidates in the main belt}

As discussed above, the existence of V-type asteroids outside the Vesta family, and in particular in orbits that are not likely to be attained by any asteroid scattered from the Vesta family, is challenging, and suggests that there were multiple basaltic asteroids in the early solar system.  
For this reason we first analyzed the orbital distribution in the space of proper elements and separated members and nonmembers of the Vesta dynamical family according to  \cite{2015aste.book..297N}.  

\begin{figure}
\begin{center}
\includegraphics[width=8cm]{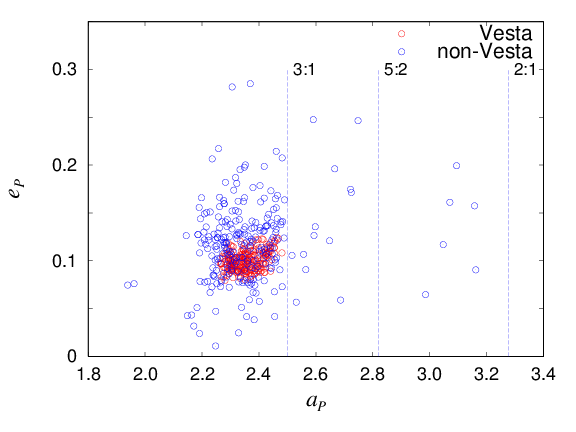}
\includegraphics[width=8cm]{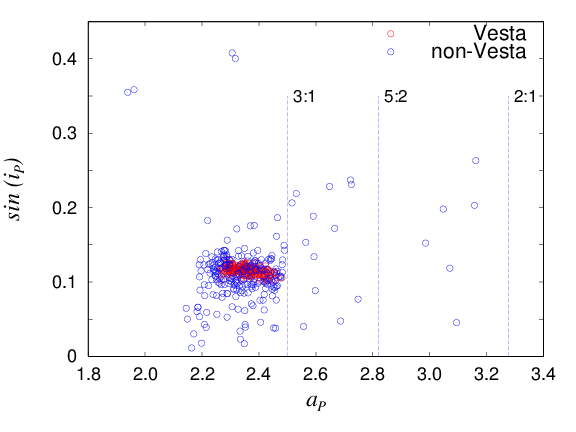}
\end{center}
\caption{Distribution in proper orbital elements of our V-type candidates. The upper panel shows the proper semi-major axis ($a_p$) vs. the proper eccentricity ($e_p$). The lower panel corresponds to the proper semi-major axis vs. the sine of the proper inclination ($i_p$). Red circles are Vesta family members, while blue circles indicate the objects not in the Vesta family. The vertical dashed lines correspond to the most relevant mean motion resonances.}
\label{VFamilyProperElements}
\end{figure}

Figure \ref{VFamilyProperElements} shows the distribution in proper orbital elements of the V-type candidates presented in Tables \ref{VVesta} and \ref{VnoVesta}. It is clearly seen that the great majority of the non-Vesta candidates are in the inner main belt (objects with semi-major axis $a_p < 2.5$ AU, hereafter IMB), most of them close to the region occupied by the Vesta family asteroids. This suggests that the large majority of those IMB V-type asteroids outside the Vesta family are probably fragments of (4) Vesta;  some of them are probably not included in the family because of the distance criterium $d_{cutoff}$ used by \cite{2015aste.book..297N} or because some of them have been scattered from the family  \citep{2008Icar..193...85N}.

On the other hand,  V-type candidates that are in the outer main belt (objects with semi-major axis between 2.8 - 3.3 AU, hereafter OMB)  and in the middle main belt (objects with semi-major axis between 2.5 - 2.8 AU, hereafter MMB) are very interesting because they are unlikely scattered Vesta family asteroids. 

Asteroids in the MMB are objects that are beyond the 3:1 mean motion resonance with Jupiter, the outer edge of the Vesta family. According to \cite{2006Icar..183..411R}, only fragments of the Vesta family ejected at very high velocities ($> 700$ m/s) are able to be injected in the MMB. We identified 13 V-type candidates in the MMB (see Table \ref{VcandNoVesta}), \cite{2008Icar..198...77M}  identified  12 more, and \cite{2006Icar..183..411R} another 2; asteroid (40521) was identified in all  three samples. A total of 27 V-type candidates are identified in the MMB either by their near-infrared (this work) or SDSS colors \citep{2008Icar..198...77M,2006Icar..183..411R}. 

According to \cite{2015aste.book..297N}  two of the V-type asteroids in the MMB identified in this work, (197480) and (180703), belong to the (173) Ino collisional family; among those identified in \cite{2008Icar..198...77M}, asteroid  (208899) belongs to the (2732) Witt family, (84021) belongs to the (170) Maria family, and (55550) belongs to the (15) Eunomia family. The case of (55550) is very interesting because the Eunomia family has been suggested as a possible source of basaltic-type asteroids \citep{2007A&A...473..967C,2008Icar..194..125R}. Twelve of the V-type candidates identified in the MMB are in the vicinity of the Eunomia family in the space of proper orbital elements.

The case of V-type asteroids in the OMB and the possible link with (1459) Magnya is crucial in order to validate the hypothesis of the presence of multiple basaltic asteroids in the early solar system. We identified six V-type candidates in the OMB (see Table \ref{VcandNoVesta}), one of which --  (126981) -- belongs to the (9506) Telramund family. \cite{2008Icar..198...77M} identified four V-types in the OMB, and \cite{2006Icar..183..411R}  identified another six  V-types.

Finally, it is interesting to study the V-type candidates in the IMB that are far from the space of proper orbital elements of the Vesta dynamical family. This can help explain what dynamical path was taken by the Vesta asteroids to reach these orbits, and can test whether there are basaltic asteroids in the IMB that are not chunks of the crust of Vesta.  In particular there are two groups: (1) those with $a_p \le 2.22$ AU that are well inside the 7:2 mean motion resonance (between 2.23 and 2.27 AU),
which is the inner edge of the Vesta family, and  (2) those with proper inclination well below the proper inclination of the Vesta family. \cite{2008Icar..193...85N}  showed that scattered Vesta family asteroids populate most of the semi-major axis extent of the IMB, but do not extend to proper inclinations much smaller than that of the family (e.g., $i_p < 3 \degr$). There are 19 V-type candidates with  $a_p \le 2.22$ AU and 16 with $i_p \le 3.0\degr$ in our sample.  
From our list of V-type candidates in the IMB 13 of them dynamically belong to the (8) Flora family, 5 belong to the (44) Nysa family, 2 belong to the (434) Hungaria, and 1 belongs to the (27) Euterpe family. 

Spectroscopic observations of V-type candidates in the MMB, the OMB, and other non-Vesta family V-types are extremely important in order to confirm their taxonomical classification as V-types, and to allow mineralogical studies to compare their surface properties to those of Vesta family asteroids (e.g., \citealt{2009P&SS...57..229D,2016MNRAS.455..584F,2016MNRAS.455.2871I}). These asteroids, together with dynamical studies to determine their origin (e.g., \citealt{2014MNRAS.439.3168C,2008Icar..193...85N,2014MNRAS.444.2985H,2008Icar..194..125R}), are crucial to understanding the population and origin of basaltic asteroids in this region of the belt.

\begin{table*} 
\caption{Near-infrared colors and proper orbital elements of the V-type candidates identified in the outer and middle main belt.}
\centering
\begin{tabular}{l l c c c c c c c}\hline\hline
 Number   & Name              & $(Y-J)$ & $(Y-J)_{err}$  &$(J-K_S)$& $(J-K_S)_{err}$    & $a_p$      & $e_p$    & $sin(i_p)$  \\ \hline 
 2452     & Lyot              & 0.548  & 0.006  &  0.216   & 0.017     & 3.157603   & 0.157434  & 0.203073 \\
 26842    & Hefele            & 0.543  & 0.019  &  0.200   & 0.054     & 3.070801   & 0.161099  & 0.11862  \\
 47837    & 2000 EB118        & 0.513  & 0.052  &  0.188   & 0.099     & 3.047409   & 0.116868  & 0.198066 \\
 112839   & 2002 QJ18         & 0.847  & 0.057  &  0.297   & 0.090     & 3.161972   & 0.090493  & 0.263182 \\
 123786   & 2001 BS17         & 0.560  & 0.081  &  0.143   & 0.096     & 3.094369   & 0.199381  & 0.045462 \\
 126981   & 2002 FW26         & 0.615  & 0.042  &  0.231   & 0.087     & 2.985409   & 0.064602  & 0.152313 \\ \hline
 6442     & Salzburg          & 0.545  & 0.032  &  0.023   & 0.078     & 2.686807   & 0.058888  & 0.047547 \\
 7459     & Gilbertofranco    & 0.788  & 0.010  &  0.125   & 0.019     & 2.597774   & 0.135546  & 0.088471 \\
 13760    & Rodriguez         & 0.558  & 0.029  &  0.142   & 0.072     & 2.556707   & 0.106708  & 0.040244 \\
 20188    & 1997 AC18         & 0.693  & 0.024  &  0.067   & 0.058     & 2.593472   & 0.12641    & 0.133965 \\
 40521    & 1999 RL95         & 0.668  & 0.034  &  0.073   & 0.075     & 2.531127   & 0.05667    & 0.218938 \\
 52132    & 5034 P-L          & 0.602  & 0.073  &  0.216   & 0.046     & 2.590746   & 0.247551  & 0.188448 \\
 59423    & 1999 GE4          & 0.520  & 0.027  &  0.270   & 0.075     & 2.648131   & 0.120942  & 0.228428 \\
 61354    & 2000 PY10         & 0.690  & 0.027  &  0.029   & 0.084     & 2.516033   & 0.105526  & 0.206352 \\
 61985    & 2000 RW30         & 0.679  & 0.024  &  0.101   & 0.045     & 2.564561   & 0.090761  & 0.153342 \\
 100772   & 1998 FN34         & 0.614  & 0.043  &  0.289   & 0.092     & 2.748188   & 0.246466  & 0.076894 \\
 180703   & 2004 HW46         & 0.502  & 0.031  &  0.141   & 0.093     & 2.721743   & 0.1747     & 0.237132 \\
 197480   & 2004 BE8          & 0.510  & 0.020  &  0.221   & 0.052     & 2.724477   & 0.171389  & 0.231051 \\
 249866   & 2001 QM165        & 0.929  & 0.077  &  0.273   & 0.097     & 2.666192   & 0.196148  & 0.172099 \\
\hline
\end{tabular}
\label{VcandNoVesta}     
\end{table*}

\subsection{Near-infrared color distribution of the V-type candidates}

\begin{figure}[h]
\begin{center}
\includegraphics[width=8cm]{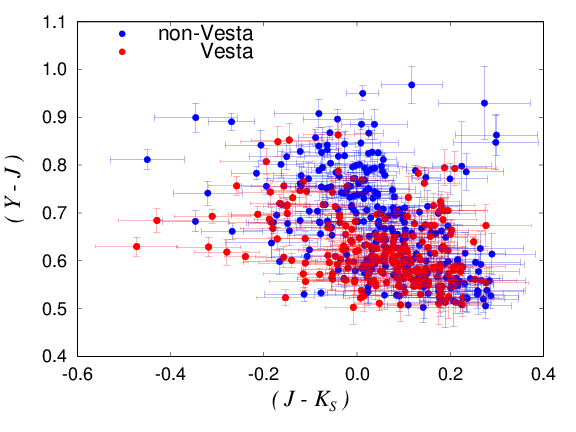}
\end{center}
\caption{VISTA colors of our V-type candidates. In red the objects belonging to the Vesta family according to \cite{2015aste.book..297N}; in blue the asteroids that are not members of the Vesta family.} 
\label{VtypeColors}
\end{figure}

\begin{figure}
\begin{center}
\includegraphics[width=8cm]{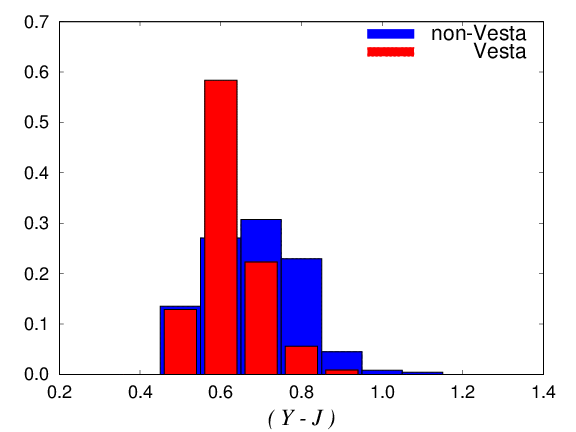}
\includegraphics[width=8cm]{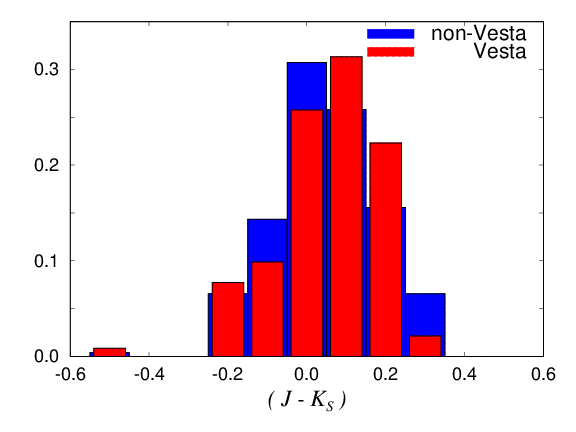}
\end{center}
\caption{Normalized histograms showing the $(Y - J)$ and $(J - K_s)$ color distributions of our V-type candidates. In red the objects belonging to the Vesta family according to \cite{2015aste.book..297N}; in blue the asteroids that are not members of the Vesta family.}
\label{VhistColors}
\end{figure}

The $(Y - J)$ vs. $(J - K_s)$ plot of the V-type candidates is presented in Figure~\ref{VtypeColors}. A first look at this figure suggests that the colors of the Vesta and non-Vesta V-types are slightly different. To better analyze this possible difference the normalized histograms showing the $(Y - J)$ and $(J - K_s)$ color distribution of both groups are presented in Figure~\ref{VhistColors}, and the  mean value and  corresponding standard deviation ($\sigma$) of the colors as well as the number of objects used to compute them (N), separated into Vesta and non-Vesta family asteroids, are presented in Table \ref{VtypeStatistical}. 

The $(Y - J)$ colors of the Vesta family candidates seem to have a narrower distribution than the colors of the non-Vesta family asteroids. The $(Y - J)$ colors of non-Vesta candidates spread over a wider range of values and peak at a larger $(Y - J)$, but the mean $(\overline{Y-J})$ values of the two distributions are similar within a 1-$\sigma$ deviation. To check if the $(Y - J)$ color distribution of the two groups are statistically different, we use a Kolmogorov-Smirnov  (K-S) test. This test assumes that both distributions are compatible and compute the probability that the two distributions are equal. The probability value $P$ obtained with the K-S test for the $(Y - J)$ distributions is $P_{(Y - J)} = 2.5 x 10^{-11}$, and for the $(J - K_s)$ is $P_{(J - K_s)}=0.24$.
To reject the null hypothesis the K-S rejection criteria range from the less strict $P < 0.05$ to the more strict $P < 0.03$ (or even $P < 0.01$), so the test cannot reject the null hypothesis in the case of the  $(J - K_s)$ distributions, but shows that the $(Y - J)$ distributions of the Vesta and non-Vesta candidates are significantly different. 



The different $(Y - J)$ color distribution of both groups can be interpreted in terms of (1) a different composition of the non-Vesta asteroids with respect to Vesta, supporting an origin in a different parent body than Vesta with a slightly different basaltic composition; (2) a different degree of space-weathering that either affects the spectral slope or the depth of the bands (see, e.g., \citealt{1993JGR....9820817P,2006Icar..184..327B}; and references therein); and (3) a different degree of contamination of the two samples by the incorrect identification of V-types using MOVIS. In this context the observed narrower distribution of the $(Y - J)$ color of the V-type candidates in the Vesta family is expected as they are chunks of the same asteroid (Vesta), ejected at the same time (thus the affected by weathering during the same amount of time in a similar region of the solar system), and the sample is likely less affected by objects that are incorrectly identified as V-type as it is biased in favor of Vestoids. 

Differences in the composition of V-types outside the Vesta family have been suggested by \cite{2016MNRAS.455.2871I} for V-types in the MMB and OMB from the analysis of four V-types outside the 3:1 resonance.  As discussed above, $(Y - J)$ is related to  Band I of the V-type spectrum and $(J - K_s)$ is related to Band II  (see Fig. \ref{Vspectrum}). The position and the depth of the bands is used to study the mineralogical composition of the V-type asteroids (see, e.g., \citealt{2005M&PS...40..445D}). In particular, the Band I center vs. Band II / Band I area ratio (BAR) is a diagnostic of different surface composition. For V-type asteroids with similar $(J - K_s)$, a larger $(Y - J)$ value can be due to either a deeper Band I (than a smaller BAR value), or a lager value of Band I center. In any case, the interpretation in terms of mineralogical composition with only these two broadband colors is not straightforward and needs to be studied further, for example by  using the spectra of basaltic meteorites, but this is beyond of the scope of this paper.

On the other hand, when considering only the OMB and MMB V-types (the non-Vesta asteroids that are more likely chunks of other basaltic parent bodies), we obtain a mean value of $(\overline{Y-J}) = 0.63 \pm 0.12$, almost the same value obtained for the Vesta family candidates (see Table \ref{VtypeStatistical}). Unfortunately, there are not enough V-types in the OMB and MMB   to perform a statistically significant $(Y - J)$ distribution to compare with the Vesta asteroids, so we cannot claim  any compositional difference between the two populations. Visible and near-infrared spectra of a significant number of these non-Vesta V-types, in particular in the OMB and MMB, is needed in order to perform a detailed mineralogical analysis and to compare their composition with that of Vesta family asteroids.

\begin{table}[h] 
\caption{Mean values of the near-infrared colors of our V-type candidates, separated into members and nonmembers of the Vesta collisional family.}
\centering
\begin{tabular}{l c c c}\hline\hline
Color & Mean & $\sigma$ & N \\ \hline
\multicolumn{4}{l}{{\em Vesta family}}\\ \hline
$(Y-J)$         & 0.62  & 0.07  & 233 \\
$(Y-H)$ & 0.75  & 0.12  & 90 \\
$(Y-Ks)$        & 0.68  & 0.12  & 232 \\
$(J-H)$         & 0.12  & 0.13  & 94 \\
$(J-Ks)$        & 0.05  & 0.13  & 233 \\
$(H-Ks)$        &-0.07  & 0.12  & 233 \\ \hline
\multicolumn{4}{l}{{\em Non-Vesta family}} \\ \hline
$(Y-J)$         & 0.68  & 0.11  & 244 \\
$(Y-H)$         & 0.78  & 0.14  & 117 \\
$(Y-Ks)$        & 0.73  & 0.14  & 244 \\
$(J-H)$         & 0.10  & 0.14  & 120 \\
$(J-Ks)$        & 0.05  & 0.13  & 244 \\
$(H-Ks)$        &-0.05  & 0.13  & 121 \\
\hline
\end{tabular}
\label{VtypeStatistical}     
\end{table}

\section{The Vesta family}

We also used MOVIS-C to study  the near-infrared colors of the Vesta dynamical family and to combine them with the analysis of visible colors from the SDSS  and visible albedos $p_V$ from the WISE data.

\begin{figure}[h]
\begin{center}
\includegraphics[width=8cm]{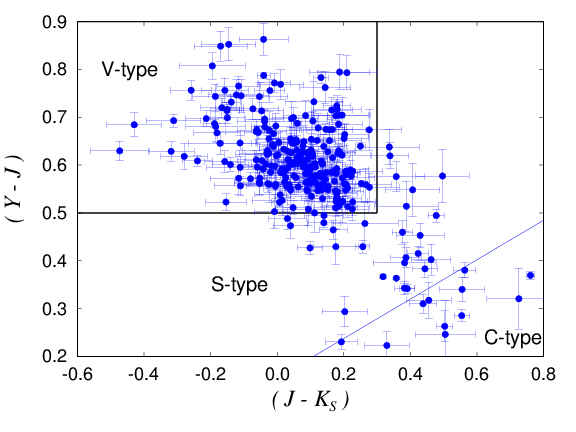}
\end{center}
\caption{ Colors of asteroids belonging to the Vesta family in the MOVIS-C catalogue with uncertainties $< 0.1$ mag. The great majority of them are in the region of the V-type asteroids as expected.}
\label{VFamily}
\end{figure}

There are 273 Vesta family asteroids according to \cite{2015aste.book..297N} observed in the MOVIS-C catalogue with $(Y - J)$ and $(J - K_s)$ colors with uncertainties $< 0.10$ magnitudes. The $(Y - J)$ vs. $(J - K_s)$ color-color plot of the asteroids belonging to the Vesta dynamical family is shown in Fig.~\ref{VFamily}. The regions between the V-, S-, and C-type classes as determined in \cite{2016A&A...591A.115P} are also shown. The Vesta family asteroids are concentrated in two clearly differentiated groups, one in the upper left (in the region of the V-type asteroids) and the other in the lower right (in the region of the S- and C-type asteroids). A total of 233 Vesta family asteroids ($\sim$ 85\%) are V-types, 39 ($\sim$ 11\%) are located in the region of S-types, and 11 ($\sim$ 4\%) in the region of C-type asteroids. 

\begin{figure}[h]
\begin{center}
\includegraphics[width=8cm]{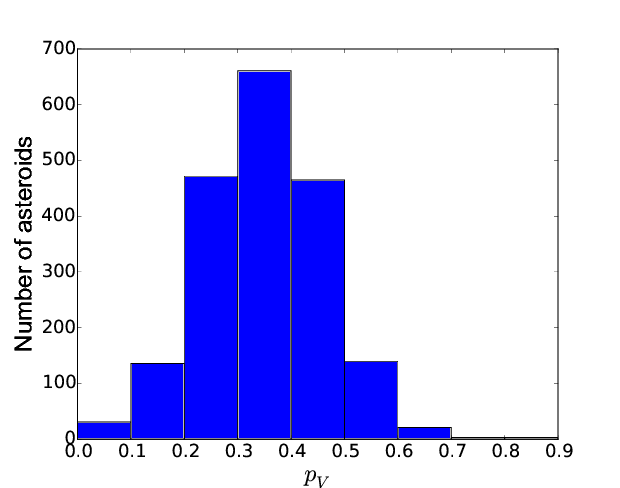}
\end{center}
\caption{ Visible albedo distribution of all the Vesta family asteroids observed by WISE. }
\label{VFamilyWISE}
\end{figure}

In this sample of 273 Vesta family asteroids 139 of them have visible albedo ($p_V$) based on WISE data \citep{2011ApJ...741...68M}. Only two of them ($\sim 1.5$ \%) have an albedo compatible with primitive asteroids ($p_V < 0.1$). The same result is obtained when considering all the Vesta family asteroids observed by WISE with determined $p_V$ (see Fig.~\ref{VFamilyWISE}).  Only 31 objects have $p_V < 0.1$ of a total of 1933, equivalent to 1.5\% of the sample. We note that the region occupied by the C-types in the $(Y - J)$ vs. $(J - K_s)$ plot also includes the X-types and the K-types. Some X-types (M- and E- types) and the K-type asteroids present $p_V > 0.1$, so the Vesta family contains a very small fraction of primitive interlopers.

\begin{figure}[h]
\begin{center}
\includegraphics[width=8cm]{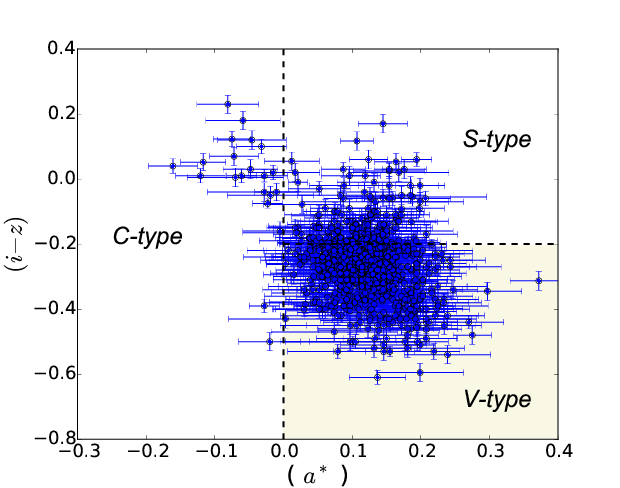}
\end{center}
\caption{ SDSS colors of the sample of Vesta family asteroids having SDSS observations. The great majority of them are in the region of the V-types according to \cite{2006Icar..183..411R}. }
\label{sdssVesta}
\end{figure}

Using the SDSS data \citep{2002AJ....124.2943I} and only considering  objects with uncertainties in the spectral slope $\sigma(a^{*}) < 0.03$ and uncertainties in the color $\sigma (i-z) <$ 0.13, there are data for 910 Vesta family asteroids (see Fig.~\ref{sdssVesta}).  A total of 659 of these objects ( $\sim 73$\%) fulfill the criteria used by \cite{2006Icar..183..411R} to determine a V-type, i.e., $(i-z) > -0.2$ and $a^{*} > 0$, and only 23 (2.5\%) are in the region of C-type primitive class asteroids. We note  that a large fraction of the asteroids in the region of the S-types are concentrated very close to the region of the V-types, 179 of them having $(i-z)$ color in the range -0.2 $< (i-z) < -0.1$).

\section{Conclusions}

In this paper we used the Moving Objects VISTA Survey Color Catalog (MOVIS-C, \citealt{2016A&A...591A.115P}), a compilation of the near-infrared colors of $\sim$ 40000 known minor bodies of the solar system observed during the VISTA-VHS survey, to search for V-type asteroids and to study the color distribution of the Vesta dynamical family. 

A total of 477 V-type candidates were identified in MOVIS-C based on their near-infrared $(Y-J)$ and $(J-Ks)$ colors, 455 of them not previously recognized by any other spectroscopic or spectrophotometric survey. This sample almost doubles the number of known V-types outside the Vesta family. The MOVIS success rate   of detecting V-types is $\sim 87 \%$, 15 objects in our list have a previous taxonomical classification based on spectroscopic observations, 13 of them were already classified as V-types. We found that 244 of the V-type candidates are not Vesta family members according to \cite{2015aste.book..297N}. We also enlarged the sample of V-types with an unlikely origin in the Vesta family:  we identified 19 V-type asteroids beyond the 3:1 mean motion resonance, 13 of which  in the mean main belt and 6 in the outer main belt, and we found other 16 in the inner main belt with proper inclination well below that of the Vesta family ($i_p \le 3.0\degr$) and well bellow the $i_p$ value that scattered Vesta family objects could attain \citep{2008Icar..193...85N}.

We also found that the $(Y - J)$ colors of the Vesta family candidates seem to have a narrower distribution compared to the colors of the non-Vesta family asteroids, and peaks at a smaller $(Y - J)$ value. \cite{2016MNRAS.455.2871I} found that V-types in the MMB and OMB seems to present a different surface composition with respect to Vesta family asteroids, but  when considering only the OMB and MMB V-types in our sample we found that their color distribution is almost equal to that of the Vesta family. Therefore, we conclude that a careful visible and near-infrared spectroscopic study of the V-types in this region of the main belt is needed in order to perform a detailed mineralogical analysis and compare their composition with that of Vesta family asteroids. 

Finally the near-infrared colors of the all the Vesta family asteroids in MOVIS-C (a total of 273 asteroids) are presented and analyzed together with their SDSS colors and visible albedo $p_V$. We conclude that the great majority of the asteroids identified as Vesta family members by pure dynamical considerations are V-types ($\sim$ 85\%), and only a few $\sim 1-2$ \% are primitive asteroids and so unlikely members of the collisional family. 

\begin{acknowledgements}
      This article is based on observations acquired with the Visible and Infrared Survey Telescope for Astronomy (VISTA). The observations were obtained as part of the VISTA Hemisphere Survey, ESO Program, 179.A-2010 (PI: McMahon). 
      We thank Daniela Lazzaro for her useful comments that helped to improve this manuscript.     
      J. Licandro, D. Morate, and J. de Le\'on acknowledge support from the AYA2015-67772-R  (MINECO, Spain). 
      The work of M. Popescu was supported by a grant of the Romanian National Authority for  Scientific  Research --  UEFISCDI,  project number PN-II-RU-TE-2014-4-2199.    
\end{acknowledgements}

\bibliographystyle{aa}
\bibliography{MPVISTA.bib}

\end{document}